\documentclass[twocolumn]{svjour3}       
\smartqed  
\usepackage[numbers]{natbib}
\usepackage{amsmath} 
\usepackage{amssymb}
\usepackage{bm}
\usepackage{graphicx}
\usepackage{gensymb}
\renewcommand{\vec}[1]{\mathbf{#1}}

\begin{document}

\title{Evolution of fragment size distributions from the crushing of granular materials}%
\author{Pavel S. Iliev         \and
        Falk K. Wittel 		   \and
        Hans J. Herrmann
}

\institute{P. Iliev \at
			  Institute for Building Materials, \\
              ETH Zurich, Switzerland \\
              Tel.: +41-44-633-74-79 \\
              \email{ilievp@ethz.ch}          
}

\maketitle
\begin{abstract}
We study the fragment size distributions after crushing of single and many particles under uniaxial compression inside a cylindrical container by means of numerical simulations. Under the assumption that breaking goes through the bulk of the particle we obtain the size distributions of fragments for both cases after large displacements. For the single particle crushing, this fragmentation mechanism produces a log-normal size distribution, which deviates from the power-law distribution of fragment sizes for the packed bed. We show that as the breaking process evolves, a power-law dependency on the displacement is present for the single grain, while for the many grains system, the distribution converges to a steady state. We further investigate the force networks and the average coordination number as a function of the particle size, which gives inside about the origin of the power-law distributions for the granular assembly under uniaxial compression.
\end{abstract}
\section{\label{sec:Introduction}Introduction }
Granular materials constitute an essential part of various natural phenomena and are present in numerous industrial processes. Their complex and counter-intuitive mechanical behavior has fascinated researchers since decades ~\cite{Jaeger1996,deGennes1999}. An important issue occurring in granular systems in which flow, compaction, tapping and vibration are present is the breaking of grains into smaller fragments. Due to its complexity, it is often disregarded in the theoretical and numerical models. When fragmentation occurs, the evolution of the particle size distribution (PSD) plays a significant role in many industrial and natural processes, like milling, sieving and segregation. Whether it is a single particle fragmentation or crushing of a granular packing under compression or shearing, there are many open questions, even though both problems have been studied extensively in the last few decades ~\cite{Steiner1974,Mehta1978,Hajratwala1982,Motzi1984,Sammis1987,Steacy1991,Astrom1998,Tsoungui1999,McDowell2001,Coop2004,Zhao2015,Bono2016}. It is a known fact that the ultimate stress of a particle is size dependent, scaling inversely to the particle size ~\cite{Vallet1995,McDowell2003,Lim2004} and the scaling law can be described by a Weibull distribution ~\cite{Weibull1939}. Because of the scaling, bigger particles tend to break more easily when they are transmitting a load through a small number of contacts. However, in a packing of particles, the bigger particles are usually surrounded by many smaller ones, leading to higher coordination numbers for the big particles, creating a state similar to hydrostatic pressure around them. Therefore, it is less likely for the bigger particles to break. The interplay of those two mechanisms leads to a power-law fragment size distribution with exponent $\alpha \approx 2.5$ for a packed granular bed under compression ~\citep{Sammis1987,Steacy1991,McDowell1996,McDowell2001,Coop2004}. The situation for the single particle crushing, however, has a different character, since the fragment interactions do not play an important role and depending on the load conditions and material properties, different size distributions are to be expected from those PSDs for the packed bed. Even though most single particle crushing experiments have been performed on quartz sand, different distributions have been observed, such as a power-law with exponent $\alpha \approx 1$ ~\cite{Yashima1970}, a superposition of log-normal distributions ~\cite{Klotz1982}, and more recently, a power-law with exponent $\alpha \approx 2$ ~\cite{Zhao2015}.

A single grain under slow uniaxial compression will have sequential failures where the particle that further breaks will be the one transmitting load between the bottom and top plates. This process leads to a gradual reduction in fragment sizes, governed by a random process. Kolmogorov ~\cite{Kolmogorov1941} and Epstein ~\cite{Epstein1948} developed the theoretical foundation for PSDs, assuming no preferential selection of fragment size which led to a log-normal distribution, which often approximates experimental observations ~\cite{Austin1939,Steiner1974,Mehta1978,Hajratwala1982,Motzi1984}. Although most materials, such as crystalline solids like quartzite sand, do have preferential crack formation regions and directions ~\cite{Zhao2015}, it is important to understand the effect of individual failure modes. Depending on the amount of input energy, particle geometry, and contact configuration, different mechanisms can be distinguished: breaking through the bulk of the particle, crumbling due to local compaction, chipping off at contact points, splitting to several pieces and disintegrating into many fragments. Since the fundamental works of {\AA}str{\"o}m \textit{et. al.} ~\cite{Astrom1998} and Tsoungui \textit{et. al.} ~\cite{Tsoungui1999} on the fragmentation of granular packings, various other models have been proposed and investigated, but the task of capturing the experimentally observed power-law distributions ~\cite{Bono2016,Zhou2014,Cantor2017c} of fragment sizes still remains challenging. The main focus is on the compression laws and the compaction behavior ~\cite{Cheng2003,Lim2005,Elias2014,Laufer2015,Hanley2015}. Concerning the single particle breaking, the most established numerical techniques employ bonded elements, such as discs(spheres) ~\cite{Potyondy2004,McDowell2002,Wang2017} or polygons(polyherda) ~\cite{Kun1996,Galindo-Torres2012,Nguyen2015,Cantor2017,Ma2017} and the focus remains on the modeling of critical breakage force and the crack propagation. Various simulation techniques have been employed to model the fragment size distributions for impact breaking of a single grain ~\cite{Kun1996,Astrom2004,Wittel2008,Timar2010} or for dynamical fragmentation ~\cite{Ma2017}, but little work has been done on the numerical modeling of PSDs at slow compression rates for large displacements, where dynamical effects can be neglected. Recently, the split-cell method ~\cite{Cantor2015} was developed and Gladkyy \textit{et. al.} ~\cite{Gladkyy2017} proposed the combined use of Mohr-Coulomb and Weibull criteria for fracture. They demonstrated good agreement for PSD with the experimental fragmentation of quartzite grain.

In this study we employ the split-cell method for grain fragmentation incorporated in the framework of the Non-Smooth Contact Dynamics (NSCD) method. NSCD has an advantage over smooth Discrete Element Method (DEM) models, since fragmentation introduces discontinuities in the moments and energies. Following Ref.~\cite{Gladkyy2017}, the critical stress is explicitly rescaled according to the Weibull distribution. We propose a novel idea to calculate the orientation of the degradation plane by a convex combination of two orientations, one based on the stress tensor, and one based on the moment of inertia tensor in order to take into account the shape of the particle and prevent unphysical cascading fragmentations. By means of this numerical model, we investigate the breaking process of a single particle under unconfined compression and the confined compression of a packed bed. Under the assumption of particle breakage that happens inside the bulk of the particle, by splitting the grain into two sub-grains when the critical stress is reached, we obtain the PSDs for both aforementioned systems. We show that the single particle fragmentation under these assumptions, neglecting other fracture mechanisms follows a log-normal distribution. Moreover, we  show that the evolution of the PSDs scales as a function of the global displacement by a data collapse of the distributions for different instances during the compression. Furthermore, we simulate the confined compression of breakable particles in an oedometric setup, starting from similar-sized grains and we show good agreement with the established power-law scaling for the size distribution of the fragments. This leads to the conclusion that the same crushing mechanisms can generate PSDs of different nature. Lastly, we obtain the evolution of the average coordination number as a function of the particle size, which shows the origin of the power-law size distributions for the compression of packed granular beds.
\section{\label{sec:level1}Numerical Model}
\subsection{\label{sec:level2}Particle interaction and motion}
The granular particles are geometrically represented as convex polyhedra defined by their vertices in both body and space fixed coordinate systems and a list of faces, containing for each face the indices of the corresponding vertices. As in Ref.~\cite{Iliev2018}, we impose disorder and asymmetry by generating randomly each particle, more precisely, the vertices of a particle are placed randomly on the surface of an ellipsoid with half-axis $a_{e} \geq b_{e} \geq c_{e}$ and a convex hull is obtained to construct the face list for the respective particle. The interaction between the particles is solved by means of the Non-Smooth Contact Dynamics (NSCD) method ~\cite{moreau1993}, which is based on volume exclusion constraint and Coulomb friction law without regularization as illustrated in Fig.~\ref{fig:signorini_coulomb}. Thus making the method particularly well suited for the modelling of dense packings of rigid, frictional particles with long lasting contacts. Because of the discontinuous nature of the contact laws (see Fig.~\ref{fig:signorini_coulomb}), for NSCD we employ an implicit scheme for the integration of the equations of motion:

\begin{figure}
  \includegraphics[width=0.5\textwidth]{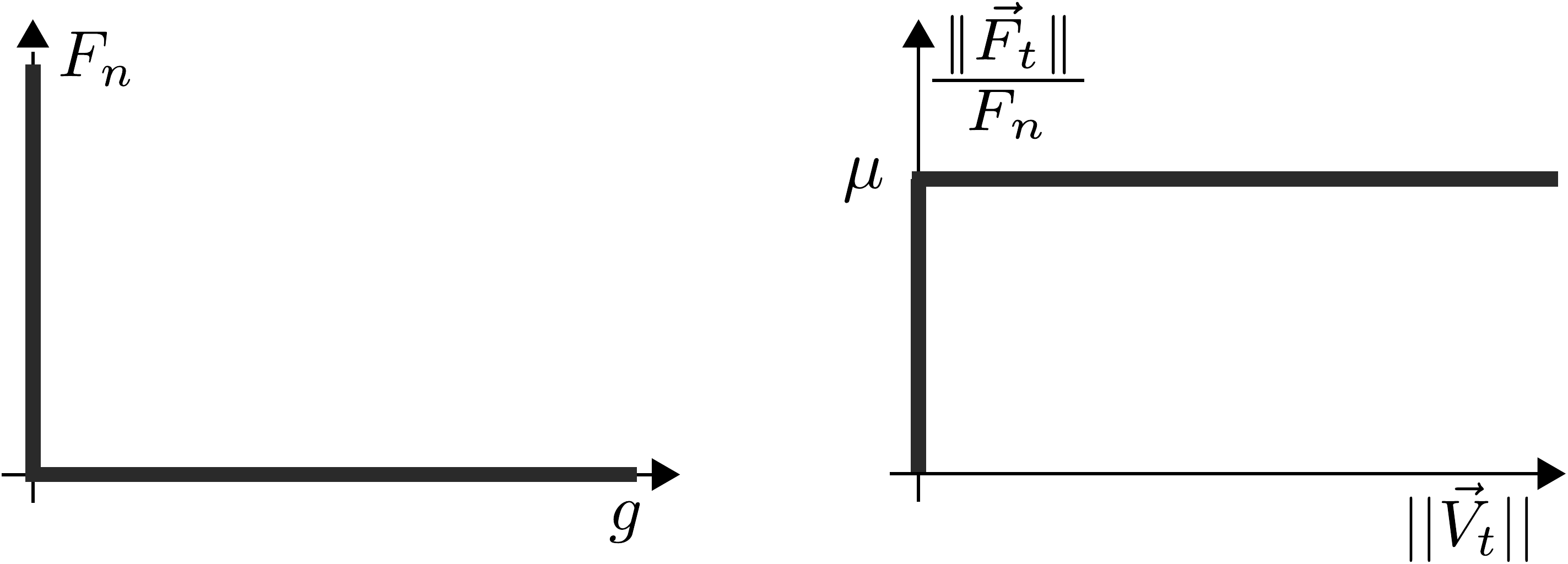}
\caption{
Contact laws for NSCD method. Left: volume exclusion constraint for the normal contact force $F_{n}$ as function of the gap $g$ between two particles (Signorini graph). Right: static friction constraint plus the dynamic friction condition expressed in terms of the relative tangential velocity $\vec{V}_{t}$, and the ratio $|| \vec{F}_{t} || / F_{t}$ of tangential and normal contact forces between two particles (Coulomb graph).
}
\label{fig:signorini_coulomb}     
\end{figure}

\begin{equation} 
\label{cd_eq}
\begin{split}
m_{i} \frac{d}{dt} \vec{v}_{i} & = \vec{F}_{i}, \\
\underline{\mathbf{I}}_{i} \frac{d}{dt} \bm{\omega}_{i} & = \vec{T}_{i},
\end{split}
\end{equation} 
where $m_{i}$ denotes the mass, $\underline{\mathbf{I}}_{i}$ the moment of inertia tensor, $\vec{v}_{i}$ the translational velocity, and $\bm{\omega}_{i}$ the rotational velocity for a particle $P_{i}$. The subscript $i$ denotes the particle number, going over all $N_{p}$ particles. $\vec{F}_{i}$ and $\vec{T}_{i}$ in Eq. ~\ref{cd_eq} are respectively the forces and torques acting on the particle. Each force $\vec{F}_{i}$ is a sum of contact forces and external forces, which we denote by $\vec{F}_{i}^{cont}$ and $\vec{F}_{i}^{ext}$. Also, the the torques $\vec{T}_{i}$ are a sum of torques due to contacts, $\vec{T}^{cont}_{i}$ and due to external sources, $\vec{T}^{ext}_{i}$. At each time step $\delta t$, the forces $\vec{F}_{i}^{cont}$ and torques $\vec{T}^{cont}_{i}$ are calculated with an iterative Gauss-Seidel algorithm until a global convergence criterion is fulfilled. The distances and the normal vectors for two contacting particles are calculated by the Common Plane (CP) method ~\cite{Cundall1988,Nezami2004,Nezami2006}. Note that for the interaction between polyhedral particles three situations may arise, namely point, line, and area contact. However, it is sufficient to represent those cases by single, double, and multiple contact points without modifying the constraint force law. 
\subsection{\label{sec:level2_1}Breaking of particles}
There are various approaches for the modelling of crushable irregular grains, such as the decomposition of aggregated particles ~\cite{Kun1996,Potyondy2004,Lim2005,Kazerani2010,Nguyen2015}, Finite Element (FE) discretization  ~\cite{Munjiza1995,Paluszny2013,Zarate2015}, and plane splitting of polyhedral particles ~\cite{Elias2014,Cantor2015,Gladkyy2017}. While the first two techniques allow for the calculation of stress fields inside a single particle, they both have the disadvantage of being computationally expensive and thus not feasible for simulations of packings composed of large number of grains. Another disadvantage of those two categories of methods is that they are strongly dependent on the subgrain resolution and the topology of the discretization. While the split plane methods don't resolve the stress distributions inside the particles, they are more computationally efficient, thus making it possible to simulate more realistic system sizes.

Our fragmentation model, motivated by the plane-splitting methods ~\cite{Elias2014,Cantor2015,Gladkyy2017} and more specifically, the recent advances proposed by Gladkyy \textit{et. al.} ~\cite{Gladkyy2017}, incorporates the Mohr-Coulomb  failure criterion with tension and compression cut-offs, degradation plane calculation taking into account both the stress state and the geometrical shape of the particle and Weibull's probabilistic theory to capture the size effect of fragmented grains.

\begin{figure}
  \includegraphics[width=0.4\textwidth]{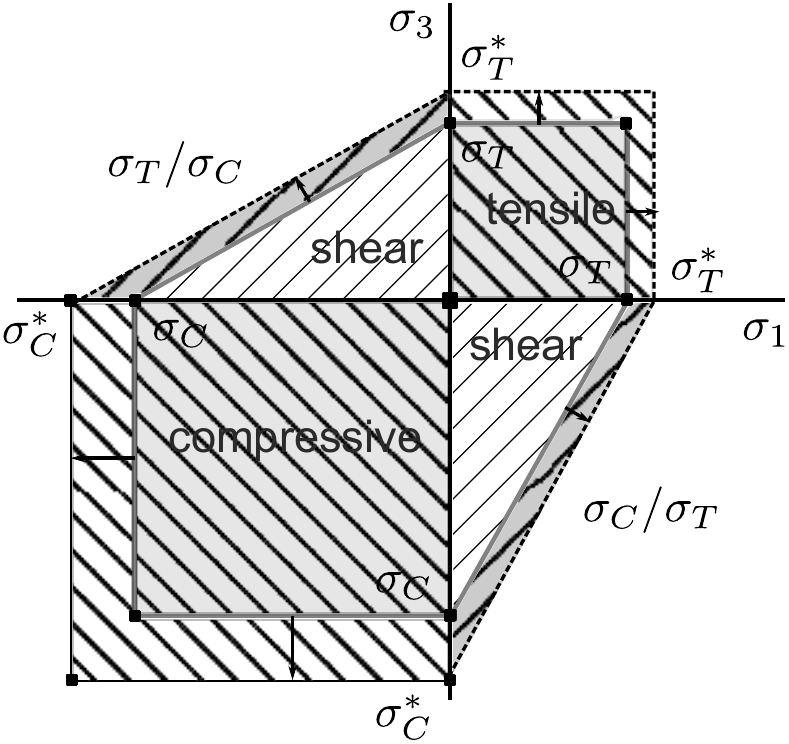}
\caption{
Mohr-Coulomb failure criterion with compression and extension cut-offs. The rescaling of the failure surface with the Weibull criterion from Eq.(~\ref{weibull_stress}) is illustrated with dashed lines and the arrows indicate the direction of the stretching as the particle diameter $d$ decreases.
}
\label{fig:mohr-coulomb}       
\end{figure}

The mean Cauchy stress tensor for a single particle $P_{i}$ can be calculated as described in Ref.~\cite{Bagi1996}:
\begin{equation} 
\label{mean_stress}
\underline{\mathbf{\sigma}}_{i} = \frac{1}{V_{i}} \sum_{c=1}^{N_{c}} \vec{l}^{(c)} \otimes \vec{F}^{(c)} ,
\end{equation} 
where $c$ spans over all $N_{c}$ contacts of the particle, and $\vec{F}^{(c)}$ and $\vec{l}^{(c)}$ are the contact force and the branch vector of the two particles forming the $c$-th contact, and $V_{i}$ is the volume of $P_{i}$. To ensure the moment equilibrium for the stress tensor $\underline{\mathbf{\sigma}}_{i}$, we perform a simple symmetrization procedure by averaging opposite non-diagonal components. After the symmetric stress tensor is constructed, the principal stresses $\sigma_{1} \geq \sigma_{2} \geq \sigma_{3}$ are calculated together with their corresponding principal axes $\vec{n}_{1}^{\sigma}$, $\vec{n}_{2}^{\sigma}$, and $\vec{n}_{3}^{\sigma}$. The Mohr-Coulomb failure criterion with cut-offs is implemented as described in Refs.~\cite{Bai2009} and ~\cite{Gladkyy2017}: the compression strength $\sigma_{C}$ and the tensile strength $\sigma_{T}$ define the limits of the failure envelope. For convenience we denote the ratio $\sigma_{T} / \sigma_{C}$ by $\sigma_{TC}$. The failure condition is composed of three primitive conditions: $\{(\sigma_{1} < 0) \land (\sigma_{3} < -\sigma_{C})  \}$,  $\{(\sigma_{3} > 0) \land (\sigma_{1} > \sigma_{T}) \}$, and $\{ \left| \sigma_{1} - \sigma_{TC}\sigma_{3} \right| > \left| \sigma_{T} \right| \}$ corresponding to compression, tensile and shear failure respectively. The failure envelope in the $\sigma_{1} - \sigma_{3}$ plane is illustrated in Fig~\ref{fig:mohr-coulomb}.

When the stress hits the failure surface at one of the primitive surfaces, the particle fragments along a degradation plane with a direction vector $\vec{n}_{p}$ passing through the center of mass of the polyhedron, resulting in two small polyhedra. In order to take into account the shape of the particle for the derivation of the splitting plane, we first calculate the principal components $I_{1} \geq I_{2} \geq I_{3}$ and the principal axes $\vec{n}_{1}^{\mathbf{I}}$, $\vec{n}_{2}^{\mathbf{I}}$, and $\vec{n}_{3}^{\mathbf{I}}$ of the moment of inertia tensor $\underline{\mathbf{I}}_{i}$. We then obtain the aspect ratios $a \leq b \leq c$ of the polyhedral particle aligned with the coordinate system defined by $ \left( \vec{n}_{1}^{\mathbf{I}}, \vec{n}_{2}^{\mathbf{I}}, \vec{n}_{3}^{\mathbf{I}} \right) $, i.e. the body fixed inertial frame of reference.

The orientation of the degradation plane is calculated according to $ \vec{n}_{p}  = \lambda^{\beta} \vec{n}_{\sigma} + \left( 1 - \lambda^{\beta} \right) \vec{n}_{ \mathbf{I} }  $, where $\vec{n}_{\sigma} = \left( \vec{n}_{1}^{\sigma} + \vec{n}_{3}^{\sigma} \right) / 2$ defines the shear plane obtained from the stress state of the particle, $\vec{n}_{\mathbf{I}} = \vec{n}_{3}^{\mathbf{I}}$ defines the axis of smallest rotational moment, and $\lambda = a/c$ is the ratio of shortest to longest aspect ratios of the polyhedron, and the exponent $\beta$ defines whether $\vec{n}_{\sigma}$ or $\vec{n}_{\mathbf{I}}$ is the dominant orientation. This expression essentially means that when $\beta > 0$, elongated particles are more likely to fragment along a plane perpendicular to their longest direction. For all simulations we use $\beta = 1$. This approach has two advantages: first, it mimics bending failure, which is not taken into account otherwise, and second, it prevents cascading fractures as discussed in Ref.~\cite{Elias2014}. After the vector $\vec{n}_{p}$ is calculated, the particle is split along the plane with orientation $\vec{n}_{p}$ passing through the center of mass of the original particle. This splitting procedure is built on the assumption that the fracture propagates through the bulk of the particle and effects like chipping and crumbling are neglected. The newly created particles are assigned the translational and angular velocities of the parent particle. 

In order to take into account the effect of particle size into the failure criterion we employ Weibull's statistical theory for the strength of materials ~\cite{Weibull1939}. According to Weibull's theory, the probability of failure $P_{f}$ as a function of the critical stress $\sigma^{c}$ is given by:
\begin{equation}
\label{weibull_prob}
P_{f} \left( \sigma^{c} \right) = 1 - \exp \left \{ - \Bigg( \frac{d}{d_{0}} \Bigg)^{3} \Bigg( \frac{\sigma^{c}}{\sigma_{0}} \Bigg)^{m} \right \} , 
\end{equation}
where $d$ is the particle diameter, $d_{0}$ is a reference diameter, $\sigma_{0}$ is a characteristic strength, and $m$ is the exponent of the Weibull probability distribution. The particle diameter $d$ is the diameter of the circumscribed sphere of each polyhedron. If we are to solve the inverse problem, i.e. for a given failure probability to derive the critical stress, we end up with the following equation:
\begin{equation}
\label{weibull_stress}
\sigma^{c} = \sigma_{0} \left \{ - \Bigg( \frac{d_{0}}{d} \Bigg)^{3} \ln \big( 1- P_{f} \big) \right \}^{m^{-1}}.
\end{equation} 
The latter expression allows us to predict the strength of a particle with a given diameter $d$ and thus rescale the failure envelope according to the particle size and the stress state that it experiences. The type of stress acting on the particle is taken into account by replacing $\sigma_{0}$ in equation ~\ref{weibull_stress} with $\sigma_{C}$, $\sigma_{T}$, or $\sigma_{S}$, depending on the type of stress - compressive, tension, or shear. The last step of rescaling the failure envelope is to calculate the effective stress $\sigma_{eff}$ on the particle for each stress type; $\sigma_{eff} = - \sigma_{3}$, $\sigma_{eff} = \sigma_{1}$, or $\sigma_{eff} = \left| \sigma_{1} - \sigma_{TC}\sigma_{3} \right|$, again corresponding to compression, tension, or shear. This way when the effective stress exceeds the critical stress, i.e. $\sigma_{eff} > \sigma^{c}$, the particle will break. This stretching of the Mohr-Coulomb failure surface is depicted in Fig.~\ref{fig:mohr-coulomb}, where the rescaled surface is shown with dashed lines and the new compression and tension strengths are denoted by $\sigma^{*}_{C}$ and $\sigma^{*}_{T}$ respectively. The shear strength is also rescaled, while the two slopes $\sigma_{T}/\sigma_{C}$ in quadrant \textit{II} and $\sigma_{C}/\sigma_{T}$ in quadrant \textit{IV} are kept constant.

\subsection{\label{sec:level2_2}Simulation procedure}
We are focusing here on the numerical simulation of compression in an oedometric test configuration with cylindrical geometry on a single particle and on a packed bed of particles. For both cases, gravity is taken into account. The cylindrical container with radius $r_{cyl}$ has a fixed bottom plate and the top plate is lowered with: constant displacement rate $\dot{u}_{z}$ starting from some initial height $h_{0}$ for the single particle case, and constant force rate $\dot{F}_{z}$ for the packed bed. The cylindrical side walls of the container are rigid and fixed, thus a confined configuration is achieved. As in Refs.~\cite{Zhou2014,Laufer2015,Bono2016}, the friction coefficient $\mu_{w}$ with the side walls is set to zero in order to minimize boundary effects. The particle-particle friction coefficient is denoted by $\mu_{p}$. For the case of single particle crushing, a particle with randomly initialized vertices is placed at the center of the bottom of the container. For the case of a packed bed, the particles, are randomly initialized with a uniform spatial distribution and uniform random orientations. Note that the ellipsoid, on which the vertices are generated is kept constant with radii $a_{e}=d$, $b_{e}=0.9d$, and $c_{e}=0.85d$ for all particles, therefore, bias from the initial PSD is removed. After the particles are initialized, they are deposited under gravity and let to relax prior to compression. During the compression, the total kinetic energy of the particles is monitored to assure that the system is in a quasi-static regime. For the confined, many particles system, fragments that have fractured more than 10 times are discarded from the simulations, similarly to Ref.~\cite{Elias2014}, since they are not contributing significantly to the force transmission and can also lead to numerical instabilities. Typical parameter values are listed in Table ~\ref{table:params}. The simulation units are made non-dimensional by choosing characteristic length $l_{c}=0.025m$, density $\rho_{c}=2500kg/m^{3}$ and acceleration $a_{c}=9.8m/s^{2}$. The characteristic time scale is then defined from the relation $t_{c} = \sqrt{l_{c} / a_{c}}$. The friction coefficient for the single particle crushing is $\mu = 0.3$ and for the confined granular packing is $\mu = 0.4$. The cylinder radius is $r^{cyl} = 0.125m$ and $r^{cyl} = 0.175m$ for the single particle and the confined bed respectively.

\begin{table}[htb]
\centering
\caption{Parameters in dimensional (physical) units and non-dimensional units used in the simulations. The compressive strength, characteristic diameter and Weibull's modulus are taken from Ref.~\cite{McDowell2003}.}
\label{table:params}
\begin{tabular}{r|l|l|l}
\multicolumn{1}{l|}{} & \begin{tabular}[c]{@{}l@{}} $\textbf{Physical}$ \\ $\textbf{Units} $\end{tabular} & \begin{tabular}[c]{@{}l@{}} $\textbf{Simulation}$ \\ $\textbf{Units} $\end{tabular} & \begin{tabular}[c]{@{}l@{}} $\textbf{Variable}$ \\ $\textbf{Name} $\end{tabular} \\ \hline
$r_{p}$                         & $0.025m$                      & $1$                & Particle radius      \\
$\rho_{p}$                      & $2500kg/m^{3}$                & $1$                & Particle density      \\
$\delta t$                      & $5 \times 10^{-5}s$			& $0.001$            & Time step      \\
$g$                             & $9.8m/s^{2}$	     	    	& $1$                & Gravity      \\
$\sigma_{C}$                    & $20MPa$                       & $32 000$           & Compressive strength      \\
$\sigma_{T}$                    & $10MPa$                       & $16 000$           & Tensile strength      \\
$P_{f}$                         & $0.6$                         & $0.6$              & Fracture probability      \\
$d_{0}$                         & $0.05m$                       & $2$                & Characteristic diameter      \\
$m$                             & $3$                           & $3$                & Weibull's modulus     \\
$\mu$                           & $0.3-0.4$                     & $0.3-0.4$          & Friction coefficient     \\
$\dot{u}_{z}$                   & $2.5 \times 10^{-4}m/s$       & $0.0005$           & Displacement rate \\
$\dot{F}_{z}$                   & $1.2kN/s$                     & $150$              & Loading rate\\
\end{tabular}
\end{table}
\section{\label{sec2:level1}Results and discussion}
\subsection{\label{sec2:level2}Unconfined breaking of a single grain}
\begin{figure}
  \includegraphics[width=0.45\textwidth]{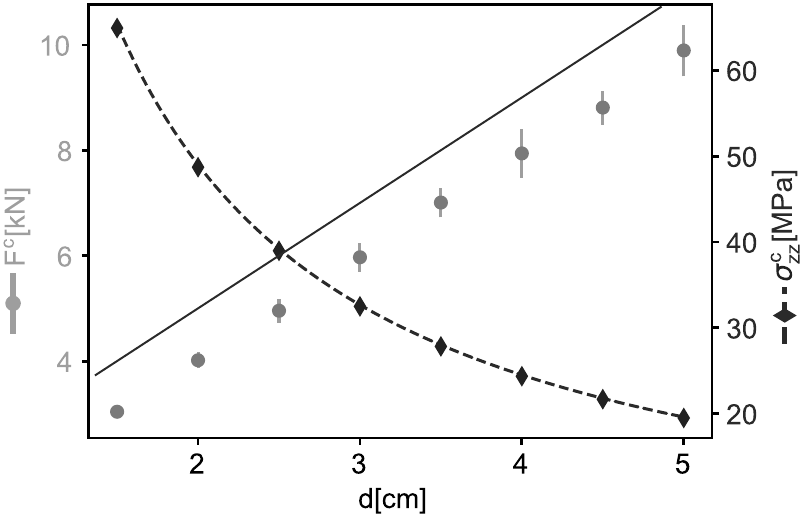}
\caption{
Critical force $F^{c}$ and critical vertical stress $\sigma^{c}_{33}$ as functions of the diameter $d$ of the generating ellipsoid(here we imply that the particles are generated from a sphere, i.e. $a_{e}=b_{e}=c_{e}$). Data points represent the mean value over 10 realizations and the errorbars represent the standard deviation. The continuous lines are the functions $F^{c}(d) = 2.0 d^{1}$ and $\sigma^{c}_{zz}(d) = 98.0 d^{-1}$}
\label{fig:crit_force_stress_vs_diameter}      
\end{figure}

We show first in Fig.~\ref{fig:crit_force_stress_vs_diameter} how the critical force $F^{c}$ experienced on the top plate and the critical stress $\sigma^{c}_{zz}$ calculated from Eq.(~\ref{mean_stress}) depend on the particle size $d$. For the chosen value of the Weibull modulus $m=3$ we define explicitly that $\sigma^{c} \propto d^{-1}$ from Eq.(~\ref{weibull_stress}). The critical force is then $F^{c} \propto d^{1}$ and as we see in Fig.~\ref{fig:crit_force_stress_vs_diameter} this relation is preserved and the fluctuations are due to the random generation of the particle.

\begin{figure}
  \includegraphics[width=0.5\textwidth]{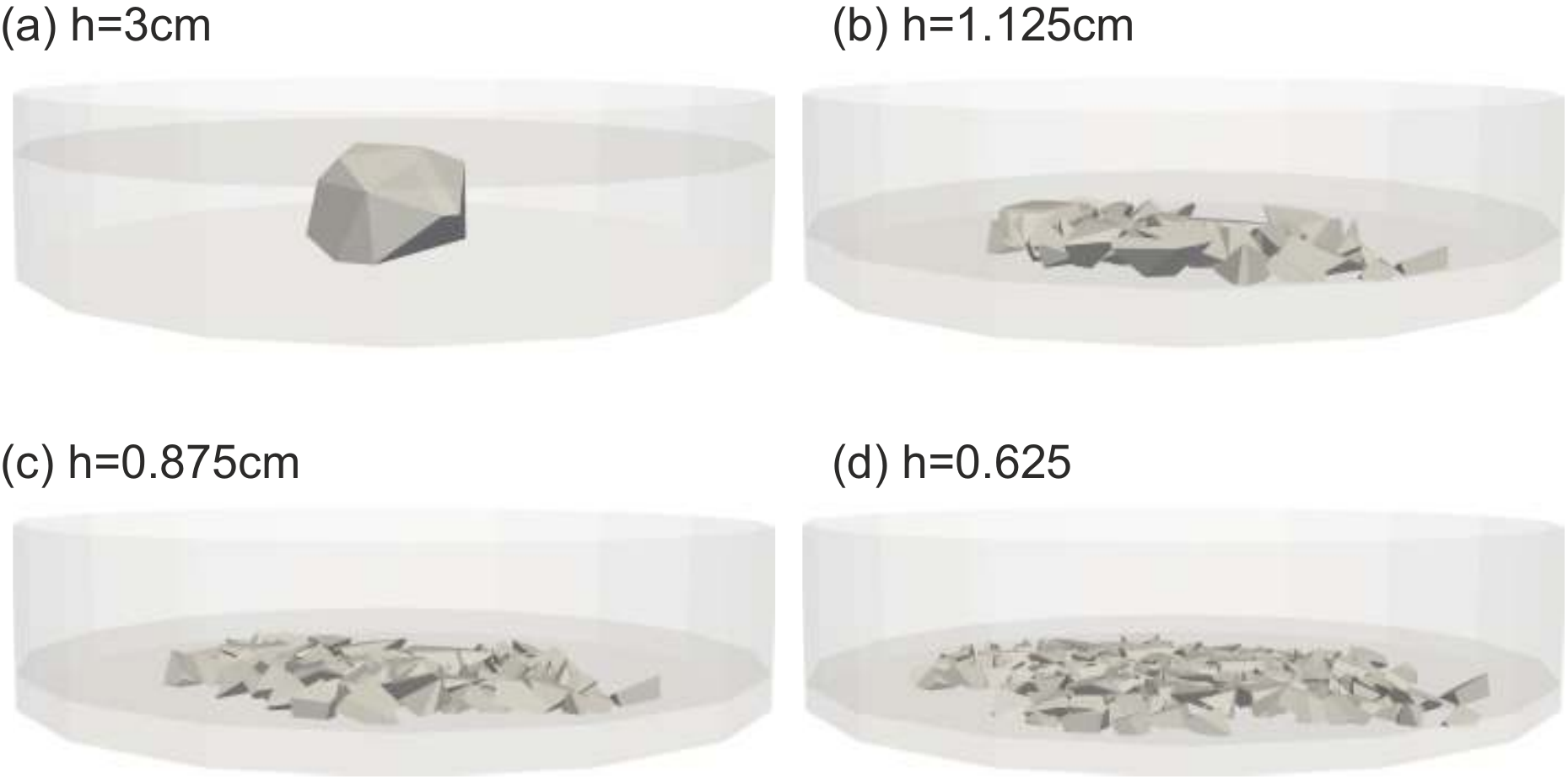}
\caption{
Snapshots from single particle compression at different heights: \textbf{(a)} $h=3cm$, \textbf{(b)} $h=1.125m$, \textbf{(c)} $h=0.875cm$, \textbf{(d)} $h=0.625cm$.}
\label{fig:1p_simulation_example}       
\end{figure}

We are interested in the mechanical behavior and size distributions after many successive fractures. Since the number of fragments from a single realization is not enough to produce robust statistical distributions, a large number of simulations has been performed to reduce the statistical noise. The initial height $h_{0}$ of the top plate of the cylindrical container is constant for all simulations in order to measure the size distributions for all realizations under the same conditions. In Fig.~\ref{fig:1p_simulation_example} snapshots of a single realization are shown for different instances. After the top plate establishes contact with the grain as in Fig.~\ref{fig:1p_simulation_example} (a), the stresses on the particle start building up until it fragments. Depending on the contact configuration and the orientation of the splitting plane, the fragments can either start sliding until they find a new stable configuration, or if the contacts remain the same for one of the subgrains, a new fracture will take place. In that case, due to the modification of the plane calculation, we observe that at most two sequential fractures can happen for the same grain. As noted previously, if $\beta = 0$, not only this effect can repeat, but the produced fragments also become very flat since the orientation of the splitting plane remains the same. After that initial breaking, if the sliding condition is satisfied for all fragments, the displacement continues with no crushing until a subgrain establishes contact with the top plate with no possibility of further rearrangements. This process is repeated many times yielding a large number of fragments when the height $h$ between the platens becomes small (see Figs.~\ref{fig:1p_simulation_example} (b), (c), and (d)).

\begin{figure}
  \includegraphics[width=0.45\textwidth]{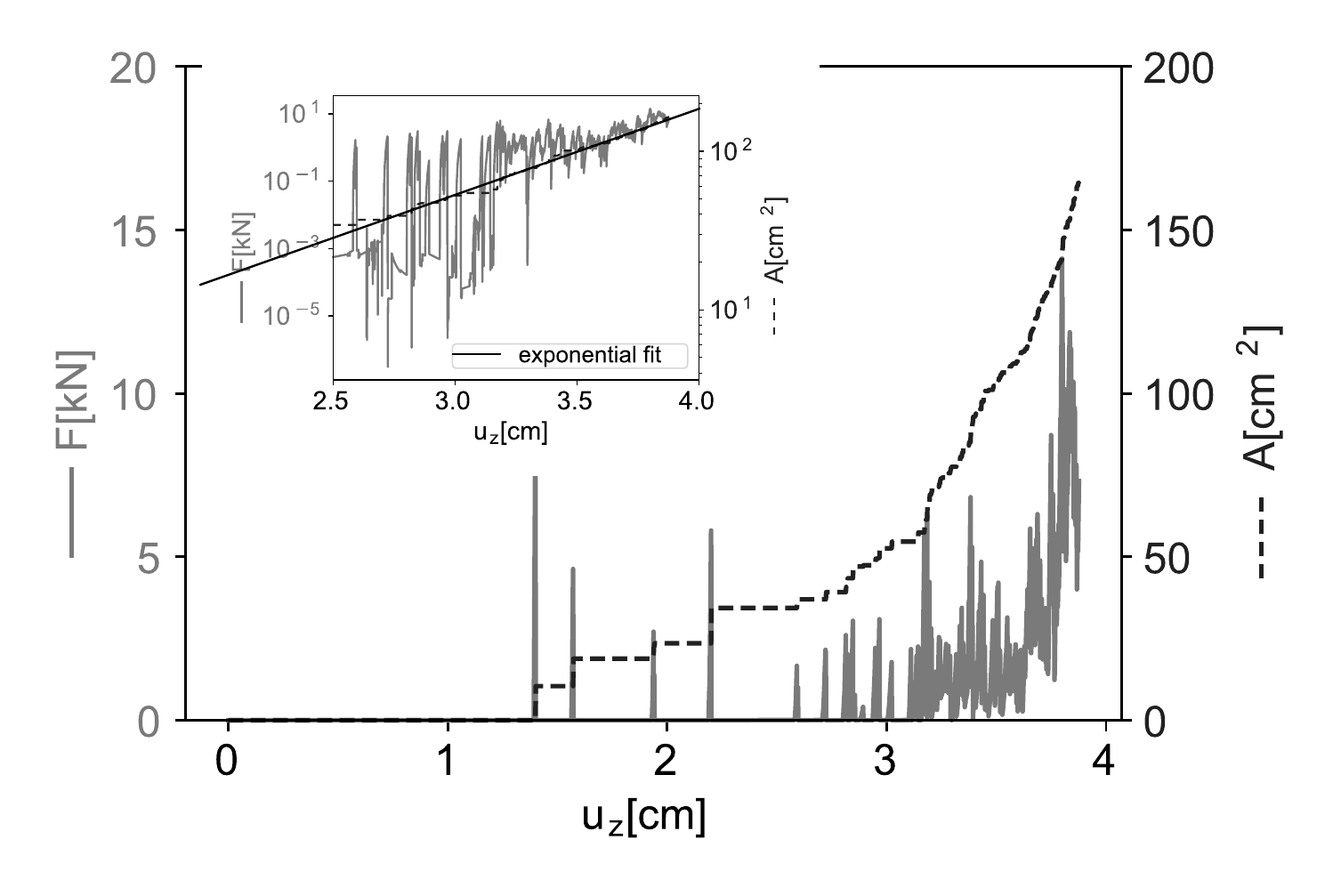}
\caption{
Resulting force $F$ on the top plate and the fracture surface $A$ from particle fragmentation as function of the top plate displacement $u_z$ for single realization. The inset shows the same graphs as in the main figure plotted in semi-logarithmic axis in the interval $u_z \in [2.5, 4]$ as well as an exponential fit of the fracture surface.}
\label{fig:force_area_vs_height} 
\end{figure}

In Fig.~\ref{fig:force_area_vs_height} we show the force $F$ at the top plate and the fracture surface $A$ generated as functions of the displacement $u_{z}$. For crack formation of brittle materials it is known that the dissipated energy is proportional to the fracture surface ~\cite{ZehnderBook}. We observe that there are initially few independent force peaks for small displacements for which the force drops to zero as the particles lose the contact with the top plate. For a larger displacement(small height $h$), i.e. $u_{z} > 3cm$ the top plate establishes contact with many fragments, leading to a collective force response at the displacing plate and $F$ does not retrace to zero. Correspondingly, the fracture surface increases with few large jumps for large displacements since the first few generations of the fragments are still of the same order as the initial particle. For large displacement, in the regime when the number of fragments is big, we see a steep exponential increase in the newly formed area. From the inset of Fig.~\ref{fig:force_area_vs_height} we see that for $u_z$ in the interval $u_z \in [2.5 cm, 4 cm]$, the fractured surface can be fitted by an exponential function. It follows that under the assumption of fragmentation through the grain bulk, the generated fracture area has an exponential dependency on the plate displacement for uniaxial compression. 

\begin{figure}
  \includegraphics[width=0.5\textwidth]{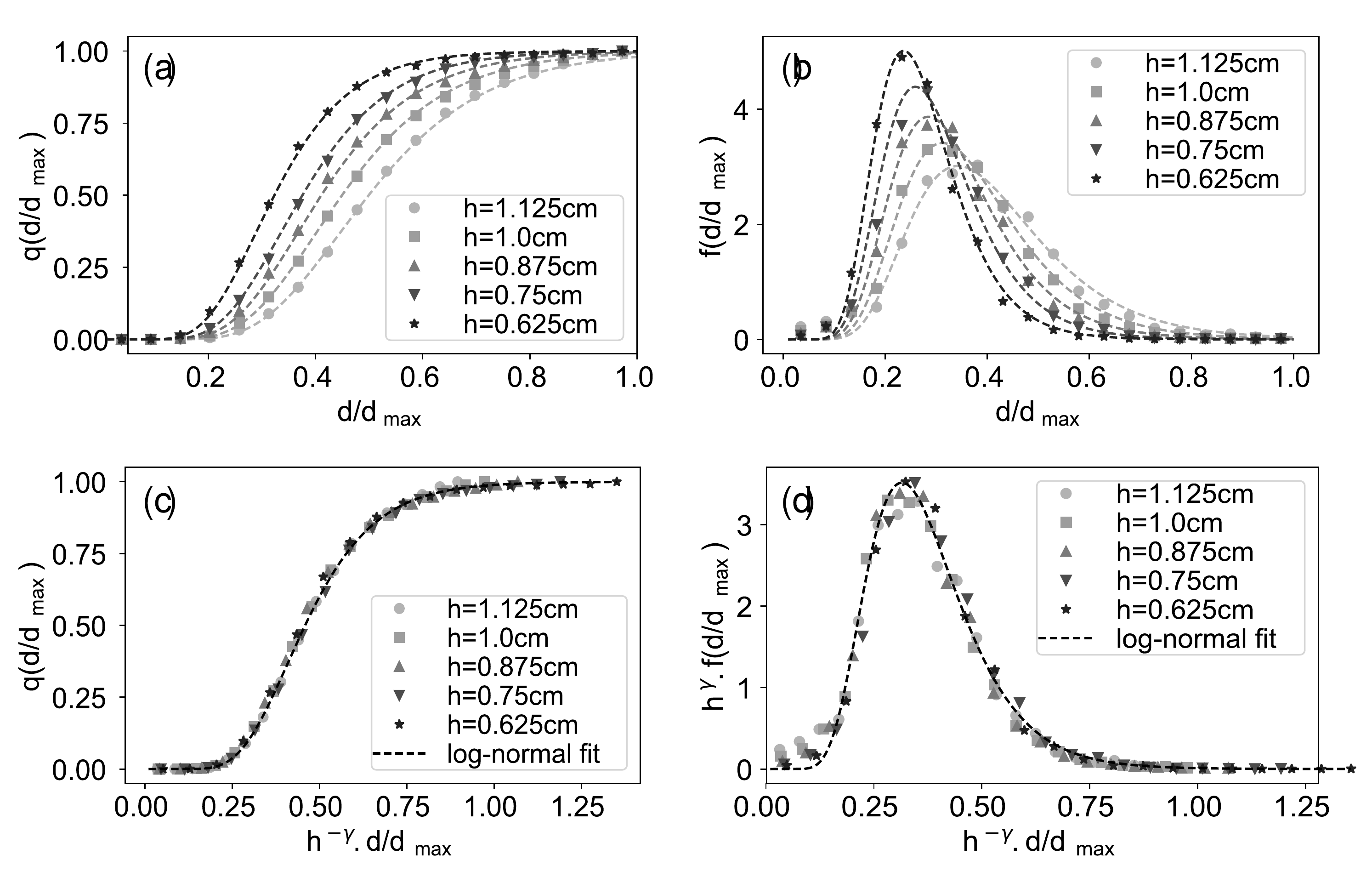}
\caption{
Fragment size distributions at different plate height $h$ averaged over 90 realizations.
\textbf{(a)} Cumulative volume fractions without rescaling.
\textbf{(b)} Probability densities without rescaling.
\textbf{(c)} Data collapse of the rescaled cumulative volume fractions.
\textbf{(d)} Data collapse of the rescaled probability densities.
Log-normal fittings are shown with dashed lines.}
\label{fig:1p_pdf_cumulat}      
\end{figure}

Next, we analyze the fragment size distributions for all performed simulations. We are interested in the probability for finding a particle with a normalized diameter $d/d_{max} \in [0,1]$. In Fig.~\ref{fig:1p_pdf_cumulat} (a), the evolution of the cumulative volume fractions (CVFs) for all realizations are shown. The distributions can be approximated by log-normal distributions with high accuracy, as seen from Fig.~\ref{fig:1p_pdf_cumulat} (a), where the distributions are fitted with log-normal cumulative distribution functions (CDFs). Since the grain size is always limited by the height $h$ of the top plate, it is expected that the largest size is fragmented as the plate is displacing. As the breaking process evolves, we see that the steepness of the distribution for $d \in [0.2 d_{max},0.6 d_{max}]$ increases, which means that the distribution is getting narrower. The dependency of the distributions as function of the plate displacement appears to be a power-law, as can be seen in Fig.~\ref{fig:1p_pdf_cumulat} (c), where a rescaling of the horizontal axis with $h^{-\gamma}$ is shown. The rescaled distributions seem to collapse nicely on a single graph for the exponent $\gamma = 0.7$.

Furthermore, we look at the probability densities (see Fig.~\ref{fig:1p_pdf_cumulat} (b)). Again, we see the pronounced shift towards the smaller sizes as well as the narrowing of the distribuions. As for the CVFs, the densities are fitted by log-normal probability density functions (PDFs) with high accuracy. Once more, we performed the rescaling of the horizontal axis by $h^{-\gamma}$, again with $\gamma = 0.7$ (see Fig.~\ref{fig:1p_pdf_cumulat} (b)). Note that since the area under the graphs has to be preserved to unity, as the plots depict PDFs, the vertical axis has to be rescaled by the inverse function $h^{\gamma}$. From the rescaled PDFs, again, a data collapse on a single graph is observed as well as a good fit with a log-normal PDF, further strengthening the assumption for a power-law dependency on the displacement. Note that some differences in the fragment size distributions between the ones obtained from our numerical model and experimentally obtained distributions for crystaline materials can occur since the formation of cleavage planes is characteristic for such materials. Other differences can be due to branching fractures, dynamical crack propagation, and the existence of other breaking modes (see Sec~\ref{sec:Introduction}) which are not captured by our model.
\subsection{\label{sec:level2_3}Confined breaking of packed granular bed}
\begin{figure}
  \includegraphics[width=0.5\textwidth]{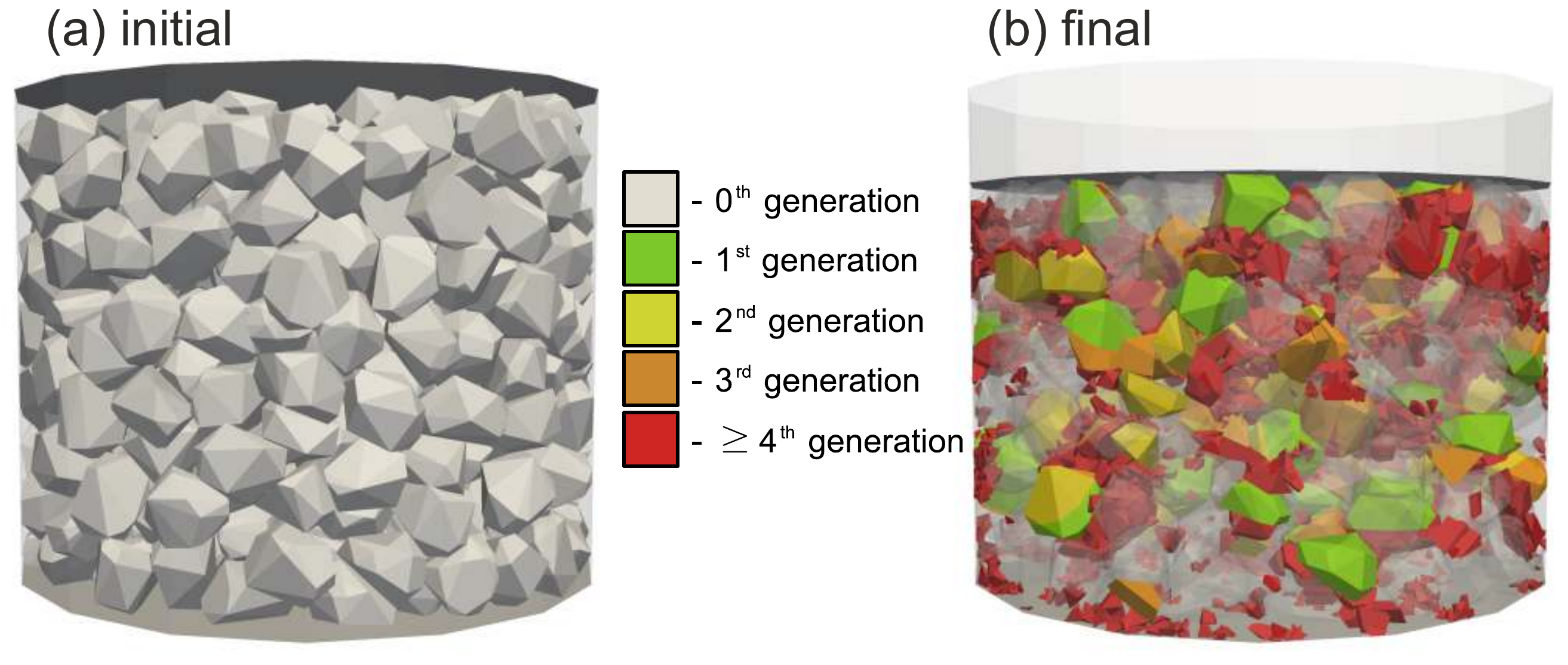}
\caption{
Snapshots of a confined packing consisting of initially $500$ particles. \textbf{(a)} Initial packing before compression. \textbf{(b)} Packing at the end of the compression at load $F=102kN$, the total number of particles is approximately $30000$. Colors represent the generation since the initial particle.}
\label{fig:Np_simulation_example}      
\end{figure}

Next, we investigate the PSDs and breaking mechanisms for a confined packed bed. The initial configuration, before loading is shown in Fig.~\ref{fig:Np_simulation_example} (a). After the compression starts, there is a regime, during which the sample is being compacted without any fragmentation, due to particle rearrangements and reconfigurations. Unlike previous numerical simulations performed with spherical particles ~\cite{Astrom1998,Tsoungui1999,Bono2016}, our model is able to capture more realistically the interlocking between individual grains and constrain their rotations. After the ultimate packing density is reached, the breaking process begins, leading to further compaction. The final state at which the simulation is stopped is shown in Fig.~\ref{fig:Np_simulation_example} (b). There are two important observations, which are crucial for the understanding of the emerging fragment size distributions. First, there is a number of grains that do not fragment even for a very large load - depicted by grey color with opacity in Fig.~\ref{fig:Np_simulation_example} (b), and second, the largest portion of fragments are the result of many breakings - depicted by red in Fig.~\ref{fig:Np_simulation_example} (b). Both of those effects are due to the same mechanism, namely that, as the system evolves, the coordination number of large particles increases significantly, leading to the decrease of the stresses that they experience. This so called ``pudding'' effect leads to the experimentally observed power-law size distributions ~\citep{Sammis1987,Steacy1991,McDowell1996,McDowell2001,Coop2004}, which we will discuss in more detail a further below.

\begin{figure}
  \includegraphics[width=0.45\textwidth]{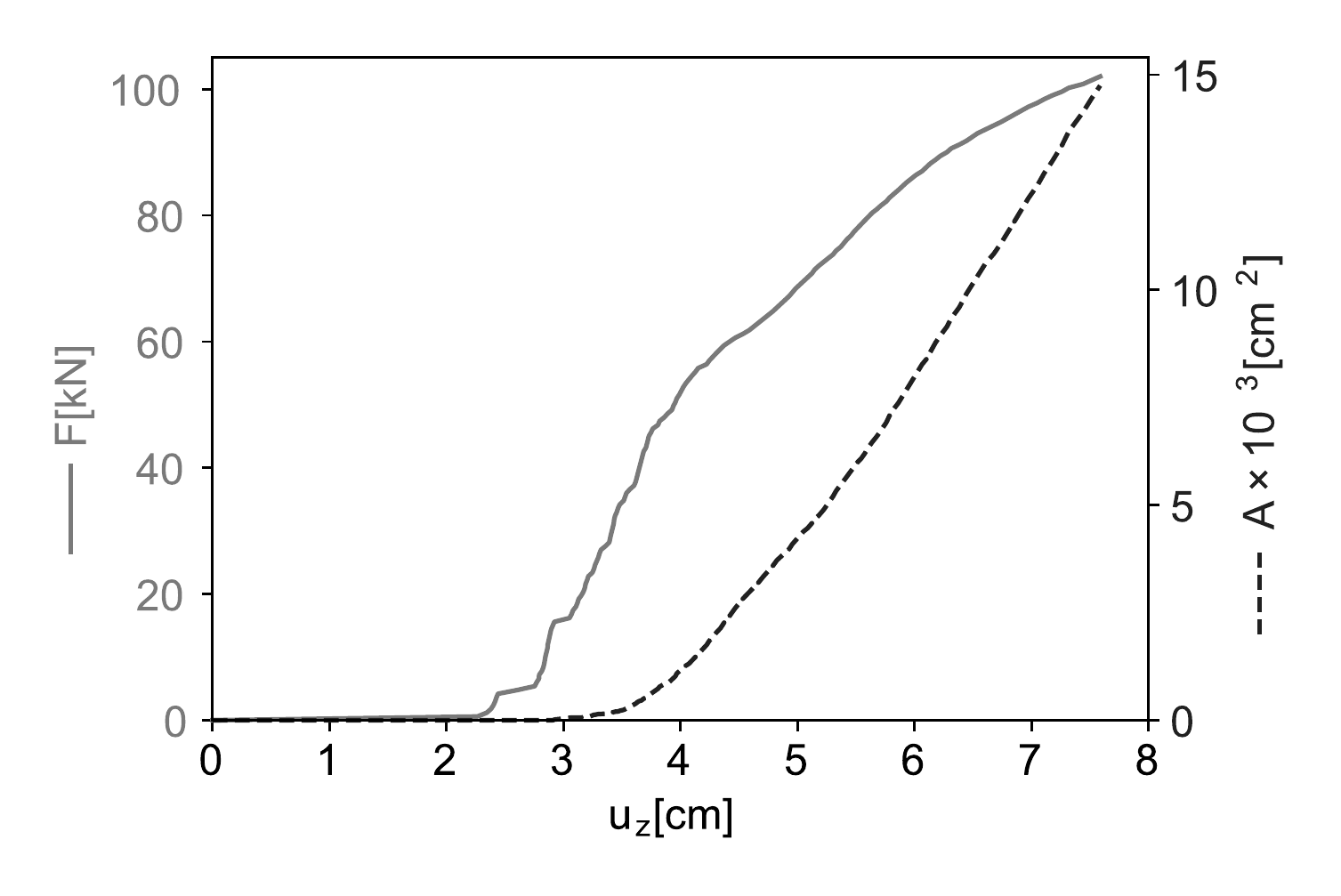}
\caption{
Applied force $F$ on the top plate and the fracture surface $A$ from the particle fragmentation as functions of the top plate displacement $u_z$. }
\label{fig:Np_force_area_vs_height}      
\end{figure}

Again, as in Sec.~\ref{sec2:level2} we analyze the force-displacement behavior as well as the fracture surface due to particle fragmentation. In Fig.~\ref{fig:Np_force_area_vs_height} we see that first the sample gets compacted without any particles getting fractured since initially the packing is loose. After the ultimate density is reached and the particles don't have enough freedom to rearrange, there is a steep increase in the applied force without significant plate displacement at $u_{z} \approx 2.6 cm$, leading to the first fractured particles at about $F=18kN$. We observe then a linear force-displacement behavior until a load of $F=60kN$ with just few large grains being broken up to this point. However, big fragments cannot fill pore spaces, thus leading to small compaction in the interval $F \in [18kN, 60kN]$. After the number of smaller fragments increases, the compaction suddenly increases, since small grains can fall on the bottom plate due to gravity or fill spaces in between large grains (see Fig~\ref{fig:Np_simulation_example} (b) for illustration). This effect allows for further compaction at a higher rate, until the simulation is stopped when the ultimate load is reached. The final displacement for this realization is $u_{z} \approx 7.6[cm]$, which corresponds to a strain $\epsilon_{z} \approx 0.25$. Interestingly, the surface area $A$ behaves very differently for the confined many particle system than for the unconfined single particle case as we see on Fig.~\ref{fig:Np_force_area_vs_height}. For the small compaction regime between $u_{z} = 2.5 cm$ and $u_{z} = 4 cm$, there is an exponential increase in the generated area due to the breaking of mostly large grains. At high compaction rate regime for $u_{z} > 5 cm$, we observe a linear dependence of the accumulated fracture area $A$ and the plate displacement $u_{z}$. This behavior is due to the emergent power-law size distribution, which will be shown bellow. 

\begin{figure}
  \includegraphics[width=0.5\textwidth]{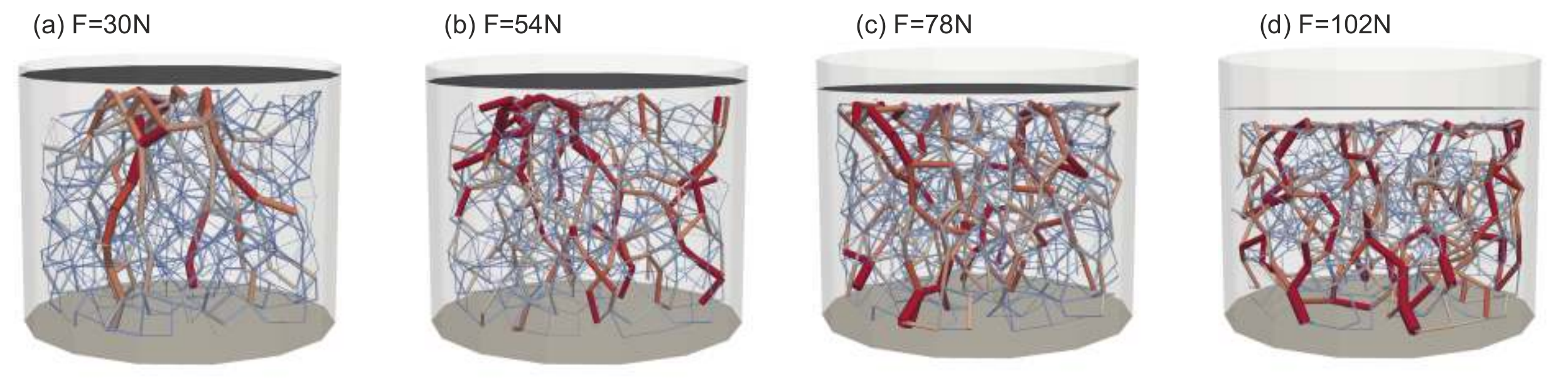}
\caption{
Force networks between grains crushed less than 4 times at different loads during the compression: \textbf{(a)} $F=30kN$, \textbf{(b)} $F=54kN$, \textbf{(c)} $F=78kN$, \textbf{(d)} $F=102kN$. The color and thickness on each segment of the network represent the magnitude of the normal contact force.}
\label{fig:Np_forcenets}    
\end{figure}

Even though, some fragments overcome the cut-off size and are removed from the simulations or end up at the bottom of the cylinder due to gravity, there is a large number of small grains preventing the percolation of vertical force chains. As we see from Fig.~\ref{fig:Np_forcenets}, at the beginning of the compression and at small loads (subfigures (a), (b), and (c)), large force chains are forming in the vertical direction, leading to the fragmentation of large grains. At the end of the compression, however, the bulk of fragmented small pieces prevents the formation of vertical force networks, which means that with very high probability the remaining large particles will not break further, even for higher loads (see Fig.~\ref{fig:Np_forcenets} (d)). 

\begin{figure}
  \includegraphics[width=0.5\textwidth]{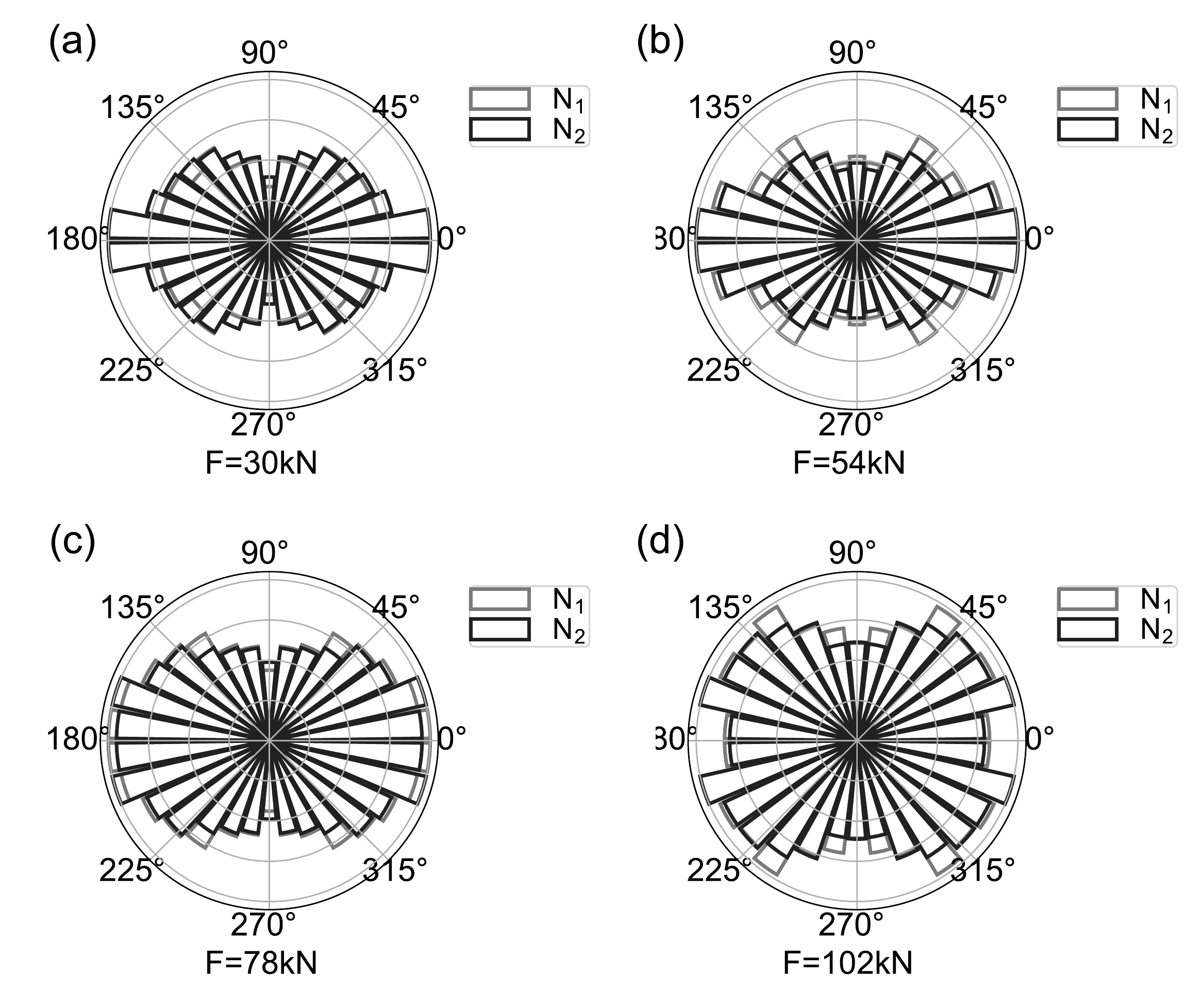}
\caption{
Normalized average normal contact force $\langle f_{n} \rangle ^{*}(\theta) $ from Eq.~\ref{force_orient_norm} of the inter particle force network as a function of the angle $\theta$ at  different loads during the compression: \textbf{(a)} $F=30kN$, \textbf{(b)} $F=54kN$, \textbf{(c)} $F=78kN$, \textbf{(d)} $F=102kN$. The data for force network N$_1$ between all particles is represented by light blue and for force network N$_2$ between all particles that are less than 4 generation is represented by light blue.}
\label{fig:Np_force_orients}     
\end{figure}

To further strengthen the hypothesis, we analyze the contact force anisotropy at different stages during the compression. We use as a measure of anistropy, the average normal contact force for a given orientation. This is done by first transforming the normal contact vector $\vec{n}$ to spherical coordinates, $(n_{x},n_{y},n_{z}) \rightarrow (n_{r},n_{\theta},n_{\phi})$, where $r=\sqrt{x^{2}+y^{2}+z^{2}}$ is the radius (since $\vec{n}$ is normalized, $r=1$), $\theta = \tan^{-1}(y/x)$ is the azimuthal angle, and $\phi = \cos^{-1}(z/r)$ is the zenith angle. Due to the axial symmetry of the system, we can neglect the influence of the azimuthal angle $\theta$. This leaves us with only one variable for the orientation of the normal vector, namely, the zenith angle $\phi$. We focus here on the average normal contact force:
\begin{equation}\label{force_orient}
\langle f_{n} \rangle (\phi) = \frac{1}{\mathopen| S(\phi) \mathclose|} \sum_{c \in S(\phi)} f_{n}^{c},
\end{equation}
where $S(\phi)$ is the set of all contacts $c$ with zenith angle $\phi$, and $\mathopen| S(\phi) \mathclose|$ denotes the size of $S(\phi)$. We obtain $\langle f_{n} \rangle (\phi)$ for two subsets of the contact force network: N$_{1}$, which is the set containing all contacts between all particles, and N$_{2}$, containing only the contacts between particles that are less than 4 generation away from the original particle. In order to compare the results for both subsets N$_{1}$ and N$_{2}$, we normalize the average force:
\begin{equation}\label{force_orient_norm}
\langle f_{n} \rangle ^{*} (\phi) = \frac{\langle f_{n} \rangle (\phi)}{\underset{\phi}{\max} ( \langle f_{n} \rangle (\phi))},
\end{equation}
and show it in Fig.~\ref{fig:Np_force_orients} for both N$_{1}$ and N$_{2}$ at different loads $F$ during the uniaxial compression. We see that for small loads (Fig.~\ref{fig:Np_force_orients} (a) and (b), when there are not many fragments, the force orientations are very similar and both are highly anisotropic, with strong peaks at $0^\circ$-$180^\circ$, indicating the dominant role of the strong force chains oriented in the vertical direction. When the load is increased and the number of fragments increases, we see for N$_{1}$, the increase of strong forces in the range of $45^\circ$-$135^\circ$ degrees as well as at the $90^\circ$ degrees. Also, the decrease of $\langle f_{n} \rangle ^{*} (\phi)$ at the $0^\circ$-$180^\circ$ degrees becomes pronounced, especially at the ultimate load. At the end of the compression, force chains are dominated by the $45^\circ$-$135^\circ$ degrees and $90^\circ$ degrees strong force orientations have significantly increased at the expense of the vertically oriented forces. The analysis of $\langle f_{n} \rangle ^{*} (\phi)$ for N$_{2}$ at higher loads shows that even if the anistropy remains in the $0^\circ$-$180^\circ$ degrees orientation, there is a significant increase of the influence of the $90^\circ$ degrees orientation. The comparative analysis of the two distributions, for N$_{1}$ and N$_{2}$, shows that the strong contacts oriented at $45^\circ$-$135^\circ$ degrees are mostly at contacts with small fragments (greater or equal to 4 fragmentation generation). Even if the anistropy of the forces between the big grains (less than 4 fragmentation generations) remains in the vertical direction, the distributions tend to become more isotropic, thus reducing the probability of a large grain to fragment.

\begin{figure}
  \includegraphics[width=0.5\textwidth]{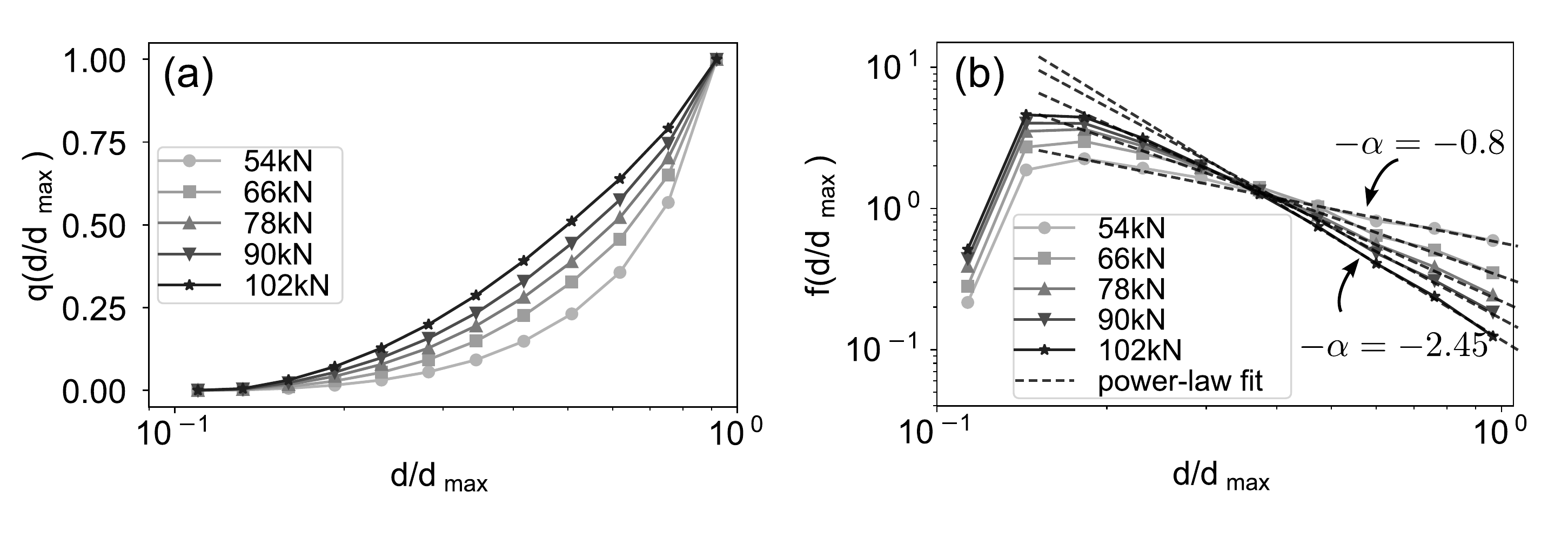}
\caption{
Fragment size distributions at different plate loads $F$ averaged over 10 realizations. \textbf{(a)} Cumulative volume fractions. \textbf{(b)} Probability densities. Power-law fittings are shown with black dashed lines.}
\label{fig:Np_pdf_cumulat}      
\end{figure}

In order to investigate in detail the behavior of the size reduction mechanisms, we analyze the fragment size distributions. Again, as in Sec.~\ref{sec2:level2}, we measure the distributions of the normalized diameter $d/d_{max}$. As seen from Fig.~\ref{fig:Np_pdf_cumulat} (a), the cumulative volume fractions are getting shifted towards the smaller sizes, appearing to converge towards a stable size distribution at the end of the compression as was previously shown in Ref.~\cite{Ben-Nur2010}. A better representation is the probability density, which is shown on Fig.~\ref{fig:Np_pdf_cumulat} (b). Note that, the smallest size introduces a cut-off of the probability density, which one can overcome with more computational resources. We see that for values of $d > 0.2d_{max}$, where the effects from the size threshold are no longer present, the distributions can be approximated with high precision by a straight line in a log-log plot. The slopes of the line fittings in a log-log scale are increasing as the system evolves. For the final size distributions at load $F = 102kN$, the slope of the fitted line is $- \alpha = -2.45$, which is very close to the established exponent $\alpha \approx 2.5$ for confined comminution ~\cite{Sammis1987,Steacy1991,Ben-Nur2010} as well as to the exponent $\alpha \approx 2.47$ of apollonian sphere packing ~\cite{Borkevic1994}. 

\begin{figure}
  \includegraphics[width=0.5\textwidth]{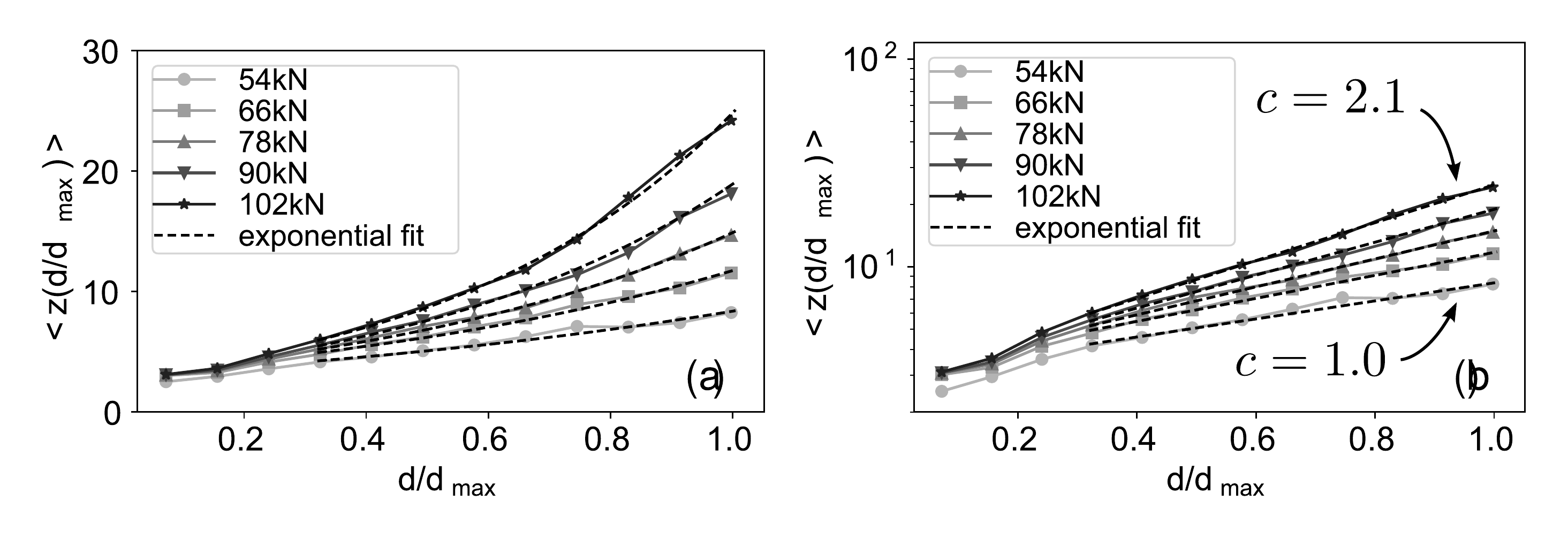}
\caption{
Average coordination $\langle z \rangle (d/d_{max})$ number as a function of the particle diameter $d/d_{max}$ at different loads $F$ averaged over 10 realizations on \textbf{(a)} linear and \textbf{(b)} semi-logarithmic plots. Exponential fittings are shown with black dashed lines.}
\label{fig:Np_cont}      
\end{figure}

Lastly, we analyze the average coordination number $\langle z \rangle$ as a function of the normalized particle diameter $d/d_{max}$. We see from Fig.~\ref{fig:Np_cont} (a) that  $\langle z \rangle$ increases as the load $F$ increases. This effect is especially strong for the large sized grains, where the average coordination number can reach mean values of up to $25$ for $F = 102kN$. One observes that the $\langle z \rangle$ is a monotonically increasing function of the particle size $d/d_{max}$ which was also shown by Bono \textit{et. al.} ~\cite{Bono2016} for the final stage of the breaking. We see that for the particle diameter $d$ in the interval $[0.3 d_{max}, 1.0 d_{max}]$, the graphs can be fitted by exponential functions, which become more pronounced as the load $F$ increases. This assumption is further strengthened by Fig.~\ref{fig:Np_cont} (b), as we plot $\langle z \rangle (d/d_{max})$ on a semi-logarithmic axis together with their exponential fittings. This leads us to the conclusion that the average coordination number has a form of an exponential function $\langle z \rangle \propto e^{c(d/d_{max})}$, where $c$ defines the slope of the linear approximation in the semi-logarithmic plot. As we see from Fig.~\ref{fig:Np_cont} (b), the slopes of those linear fits is increasing with increasing force $F$, indicating that the exponential multiplier $c=c(F)$ is a monotonically increasing function of $F$, which interestingly, does not appear to be saturating. Moreover, we obtain that with a good accuracy $c \approx $ $1.0$, $1.275$, $1.55$, $1.825$, and $2.1$ at loads $F = $ $54$, $66$, $78$, $90$, and $102$ $kN$, respectively. This leads us to the conclusion that $c$ has a linear dependency on the load $F$. This result can be explained by the fact that the the small grains increase in numbers faster than the big grains even after the stationary distribution has been reached. Therefore, the average number of contacts is increasing for the large grains and does not change much for the small grains.
\section{\label{sec3:level1}Conclusion and outlook}
We have analyzed and compared the fragment size distributions for both unconfined single particle crushing at slow compression rates and confined compression of many particles under an increasing vertical load. By means of a variation of the plane-splitting method incorporated in the framework of the NSCD method, we performed numerous simulations in order to obtain the cumulative distributions and the probability densities for both aforementioned cases. Moreover, we investigate in detail the mechanisms which cause the differences in the distributions, given the same breaking law. Since the fracture criterion is calculated based on the mean Cauchy stress for each particle, no calibration is needed to implement the correct strength scaling as a function of the particle size. Another advantage of the used method is the ability to use irregular shape representations for the grains, unlike the commonly used sphere replacement methods. This allows us to accurately model the geometrical interlocking between individual grains, which has a significant effect, especially for the packed bed system.

The breaking mechanism is build upon the assumption that the fracture propagates through the bulk of the particle and other breaking effects are neglected. Under this assumption, we obtain a log-normal fragment size distribution for the single particle crushing, which can be explained by the sequential fragmentation theory developed by Kolmogorov ~\cite{Kolmogorov1941}. Moreover, we show that there is a power-law dependency on the displacement by collapsing the data for both the cumulative distributions and the probability densities. The differences in the experimentally obtained distributions for single particle breaking can be explained by the peculiarity in the crystalline structure of the used materials, leading to predefined failure planes.

For the confined crushing of a packed granular bed, we show that unlike the single particle crushing, the fragment size distribution converges towards a stable distribution as the loading increases. The final distribution has a well defined power-law tail for particles with diameter $d$ larger than $0.2 d_{max}$ with an exponent $\alpha \approx 2.45$ which is within the range of the theoretically and experimentally obtained exponent $\alpha \approx 2.5$ ~\citep{Sammis1987,Steacy1991,McDowell1996,McDowell2001,Coop2004}. By looking at the force networks, we observe that at large loads, at which the power-law distribution is established, there are no strong vertical force chains connecting larger grains. This is indicative of the driving mechanism of the power-law size distributions, namely, the accumulation of small fragments, which redistribute the forces from the big fragments, thus reducing their stresses. This was also shown by analyzing the evolution of the normal contact force anisotropy for the force network connecting all particles as well as the contact network connecting only big fragments. Furthermore, we measured the evolution of the average coordination number $\langle z \rangle$ as a function of the particle size during the loading. We find that $\langle z \rangle$ increases for all sizes throughout the compression, but that this increase becomes steeper for bigger particles, reaching values of up to $25$ at the end of the simulation. By analyzing the results for $\langle z \rangle (d/d_{max})$, we suggest that this dependency is exponential of the form $\langle z \rangle \propto e^{c(d/d_{max})}$, where $c$ increases as the load $F$ increases. 

As an outlook for future studies, the breaking rule can be modified in order to take into account other mechanisms and incorporates a predefined degradation planes. As a first suggestion, one can take into account the contact points and define the splitting plane as a function of the weighted linear combination of the vector of the normal contact forces, as well as calculate the point at which the plane passes through the force center, instead of the mass center. Also, a comparison of the size distributions of an unconfined packed granular bed (i.e. triaxial configuration) would be an interesting topic of further investigations. Another question that can be further addressed is whether introducing tapping or shaking of the granular bed would affect the evolution of the size distributions or the average coordination number as a function of the particle size. 
\begin{acknowledgements}
We acknowledge financial support from the ETH Research Grant ``Robotic Fabrication of Jammed Architectural Structures'' ETHIIRA Grant No. ETH-04 14-2 as well as from the ERC Advanced grant number FP7-319968 FlowCCS of the European Research Council. 
\end{acknowledgements}

\end{document}